\documentclass[twocolumn]{article}
\usepackage{amsfonts,a4wide,graphicx}

\begin{document}

\title{\textbf{Broken Relativistic Symmetry Groups, Toroidal Moments and
Superconductivity in Magnetoelectric Crystals}}

\author{Jacques L. RUBIN\thanks{E-mail: jacques.rubin@inln.cnrs.fr}\\
\textit{Institut du Non-Lin\'eaire de Nice (INLN),\/}\\
\textit{UMR 129 CNRS-Universit\'e de Nice - Sophia Antipolis,\/}\\
\textit{1361 route des lucioles, 06560 Valbonne, France\/}}

\date{Enhanced and english corrected unpublished 2002 version of the 1993
one,\\ the latter being published in\\
\textit{Il Nuovo Cimento D - Cond. Matt. At.},
\textbf{15(1)} (1993), p. 59.\\
\mbox{}\\
\today}

\maketitle

\begin{abstract}
A connection between  creation of toroidal
 moments and  breaking of the relativistic crystalline group associated to
a given crystal, is presented  in this paper. Indeed, if magnetoelectric
effects exist, the interaction between electrons and elementary magnetic
cells appears in such a way that the resulting local  polarization and
magnetization break the local relativistic crystalline symmetry.
Therefore,  Goldstone bosons, also associated to  toroidal moments, are
created and, as a consequence, corresponding toroidal phases in crystals.
The list of the Shubnikov groups compatible with this kind of phases is
given and possible consequences in superconductor theory in magnetoelectric
crystals are examined.
\end{abstract}

\bigskip

\noindent\textsc{Key Words:}  \textit{tor\-oidal mo\-ments,
relati\-vistic crystal\-line sym\-metries, sym\-metry brea\-king, anyons.\/}

\bigskip

\noindent\textsc{PACS 1990:} \textit{05.30.L -- Anyons and parastatistics. 
61.50.E -- Crystal symmetry, models and space groups,
crystalline systems and classes. 74.30 -- Superconductivity, General
properties. 75.20 -- Diamagnetism and paramagnetism. 75.80 -- 
Magnetomechanical and magnetoelectric effects,
magnetostriction.\/}

\section{\normalsize\hskip-.5cm - Introduction}

The aim of this paper is to present a possible link between {\it
magnetoelectric effects} and {\it toroidal moments} via the concept of
{\it  relativistic crystalline groups} \cite{1,2,3,4,5} and to see how this
link can be applied to derive a possible origin of superconductivity in
electric and magnetic crystals. These crystals are characterized by their
magnetic groups which are subgroups of the {\it Shubnikov group} $O(3)1'$.
Among the 122 magnetic groups, only 106 are compatible with the existence
of a linear or quadratic magnetoelectric effect \cite{6}. Let us recall
that, in these groups, the time inversion $1'$ appears in addition to or in
combination with orthogonal transformations of the Euclidean space.
\par
In the present paper, we will consider relativistic symmetry group theory
in crystals. Therefore, we need an extension from the Shubnikov
 group $O(3)1'$ to the group $O(1,3)$ in the Minkowski space. More
particularly, our attention will be devoted to transformations of $O(1,3)$
leaving invariant polarization and magnetization vectors and generating a 
subgroup of the relativistic point group associated to the magnetic group
$G$ of a given crystal, namely  the normalizer $N(G)$ of $G$ in $O(1,3)$.
\par
This subgroup may not be identified to the magnetic group if $G$ leaves
invariant a particular  non-zero velocity vector, {\it i.e.}, if $G'$ and
$G\ \subseteq G'$ are isotropy groups of $O(1,3)$. If such a vector exists
and $G'\not=G$, one strictly speaks about the {\it relativistic crystalline
symmetry} $G'$. We can represent, as in figure 1, the list of the magnetic
groups for which  one can have in a crystal  a spontaneous magnetization,
 a spontaneous polarization or an invariant non-zero velocity vector
\cite{15}. Thus, as it can be seen in the figure 1, only 31 magnetic groups
are compatible with the existence of a relativistic crystalline symmetry
(see  the groups contained into the lowest circle).
The invariant non-zero velocity vectors can be linked
with toroidal moments from the point of view of magnetic symmetries as it
has been already shown in previous papers \cite{7,8}. In that one, we will
explain how toroidal moments might be generated from broken relativistic
crystalline symmetries and give a few assumptions in order to construct a
superconductivity model.

\section{\normalsize\hskip-.5cm - Preliminaries}

The toroidal moments $\bf{T}^{(\ell)}$ are polar tensors which change
sign under time inversion, like velocity vectors $\vec{v}$ or current
vectors $\vec{j}$ of electric charges.  There are referred in toroidal phase
transitions  as the {\it  order parameter} and in particularly, in the
superdiamagnetism of superconductors or dielectric diamagnetic bodies
containing densely packed atoms (\textit{``agregates"\/}) \cite{7,8,9,10,11}. In such systems, in presence of spontaneous currents,
there may exist states for which the configuration of the associated
currents has a {\it tore-shaped  solenoid with a winding}
and so a dipolar toroidal moment (see figure 2).\par
More generally, the toroidal moments appear as  form factors in the
multipolar Taylor  expansion of the current vectors satisfying the
four-dimensional conservation law
\[{\partial_{\mu}}j^{\mu}(\vec{x},t) = 0\, .\]
For a system with a dipolar toroidal moment $\vec{T}$, the current density
equals
\[\vec{j}= curl(curl(\vec{T}))\,.\]
The vector $\vec{T}$ is equal itself to $\vec{x}\,{\xi}\,(\vec{x},t)$,
where ${\xi}(\vec{x},t)$ is the second Helmotz  potential. In case of
static  multipolar electric moments, this latter  can be expanded  up to
a constant factor as \cite{12,13}  
\[\xi \equiv \sum_{k,\ell,m}k^{2}\,
{\bf F}^{\ell}_{k,m}(\vec{x})\,
{\bf T}^{\ell}_m(-k^{2},t)\, ,\]
where the ${\bf F}^{\ell}_{k,m}(\vec{x})$ functions are the basis 
functions of the Helmotz equation.
The toroidal moments components
${\bf T}^{\ell}_m(-k^{2}=0,t)$ are defined
by
\[\begin{array}{l}
{\bf T}^{\ell}_m(0,t) = 
-\frac{\sqrt{{\ell}\pi}}{(2{\ell}+1)}\,
{\huge\displaystyle\int} r^{{\ell}+1} 
\Big[
\Big\{ 
{\bf Y}^{({\ell}-1)\ast} 
\\
\left.\left.\qquad\quad
+ \frac{2\sqrt{{\ell}}}{(2{\ell}+3)\sqrt{{\ell}+1}}\,
{\bf Y}^{({\ell}+1)\ast}
\right] 
\otimes
{\bf j}^1(\vec{x},t)
\right\}^{\ell}_m\, d^{3}\!x \, ,
\end{array}\]
where $r={\parallel}{\vec x}{\parallel}$, ${\bf Y}^{({\ell})}$ is the
well-known spherical harmonic of order
${\ell}$,
${\bf j}^1\equiv{\vec j}$ is the current vector expressed in the
spherical basis and $\otimes$ indicates the tensor product of two
irreducible tensor operators. For ${\ell}=1$, we obtain the toroidal dipole
moment of a given configuration of currents, and ${\bf T}^1_m$ can be
expressed in a cartesian basis as follows
\[{\bf T}^{1}\equiv{\vec T} = \frac{1}{10} {\huge\displaystyle\int} \left\{
\vec{x}\,(\vec{x}\,.\,\vec{j})-r^{ 2}\,\vec{j}\right\}\, d^{3}\!x\, .\]
It has been shown \cite{7,8,9,10,11} that, without any other modifications,
one can formally substitute ${\vec T}$ for ${\vec v}$ in figure 1. We can
remark that this kind of configuration of currents can be created from a
solenoidal configuration of currents. The closure of the latter one can be
obtain applying an external magnetic field, but we will present another
possibility.\par
Hence, the problem is especially to determine the origin of, from one hand,
the permanent solenoidal currents and, on the other, the permanent closure
of these latter. The occurence of such phenomena is precisely discussed in
connection with the breaking of the relativistic crystalline symmetries.

\section{\normalsize\hskip-.5cm - Magnetoelectric Effects and Relativistic
Symmetries}

The state of a crystal is described with a one-valued function $f$ depending
on the electric, magnetic or/and constraint fields. Generally, $f$ is
taken to be the free enthalpy. The various properties of the crystal manifest
themselves in the appearance of various tensors as coefficients, in the Taylor
expansion of the function $f$, in terms of the different fields \cite{2,6}:
\[\begin{array}{l}
f = {}^o\!P_i\,E^i + {}^o\!M_i\,H^i\\ 
\qquad\qquad+ \frac{1}{2}
\left(\epsilon_o\,
\epsilon_{ik}\,E^i\,E^k + \mu_o\,\mu_{ik}\,H^i\,H^k\right)\\ 
\qquad\qquad\qquad+ \frac{1}{c}\,\alpha_o\,\alpha_{ik}\,E^i\,H^k +
\cdots\,,
\end{array}\]
where ${}^o\!P$ and ${}^o\!M$ are respectively the electric and
magnetic spontaneous polarizations, $\epsilon_{ik}$ and $\mu_{ik}$, the
electric permitivity  and magnetic permeability, and $\alpha_{ik}$,
the magnetoelectric succeptibility. Let us define the electric polarization 
component $P_i$ and the magnetic polarization component $M_i$ {\it via}
\[
P_i\equiv\frac{\partial f}{\partial E^i}=
{}^o\!P_i + \epsilon_o\epsilon_{ik}\,E^k +
\frac{1}{c}\,\alpha_o\alpha_{ik}\,H^k + \cdots\,,
\]
and
\[
M_i\equiv\frac{\partial f}{\partial H^i}=
{}^o\!M_i + \mu_o\mu_{ik}\,H^k +
\frac{1}{c}\,\alpha_o\alpha_{ik}\,E^k + \cdots\,.
\]
A tensor is different from zero if it remains invariant under all the
symmetry transformations of the magnetic group of the crystal. The
relativistic invariance of the non-zero permitted tensors  of the Taylor
expension has to be imposed because of the relativistic covariance of the
Maxwell equations determining the dynamics of the electric and magnetic
fields. Hence, a supplementary invariance is considered in addition to all
the previous ones. Consequently, one applies the relativistic invariance to
the second order term of the expension of $f$. It can be written \cite{4} in
the following form 
\[f^{(2)} = \frac{1}{8c}
\chi^{(\alpha\beta)(\gamma\delta)}
\,{\bf F}_{\alpha\beta}
\,{\bf F}_{\gamma\delta}\,,\]
where ${\bf F}$ is the Faraday tensor of the applied
electromagnetic field and $\chi$ the relativistic
succeptibility tensor.
\par
We can represent the pairs of indices $(\alpha,\beta)$ and
$(\gamma,\delta)$ by a single index
\[
\left\{
\begin{array}{l}
(0,1)\, \longrightarrow 1\, ,\\
(0,2)\, \longrightarrow 2\, ,\\
(0,3)\, \longrightarrow 3\, ,\\ 
(2,3)\, \longrightarrow 4\, ,\\
(3,1)\, \longrightarrow 5\, ,\\
(1,2)\, \longrightarrow 6\, .\\
\end{array}
\right.
\]
Then, $\chi$ can be represented by a $6\times 6$ matrix:
\[
\chi^{(\alpha\beta)(\gamma\delta)}
\equiv
\sqrt{\frac{\epsilon_o}{\mu_o}}
\left(
\begin{array}{rr}
-\epsilon^B&\alpha\,\mu^{-1}\\
{}^t\!(\alpha\,\mu^{-1})&-\xi\,\mu^{-1}
\end{array}
\right)
\, ,
\]
with $\epsilon^B$ = $\epsilon-\alpha\,\mu^{-1}\,{}^t\!\alpha$ ($t$ as
transposition) and where $\epsilon$, $\mu$, $\alpha$ and $\xi$ are
respectively the electric succeptibility tensor, the magnetic permeability
tensor, the magnetoelectric succeptibility tensor and the magnetic
succeptibility tensor. 
\par
The relativistic invariance imposes constraints ({\it i.e.} algebraic
expressions) between the components of the tensors $\epsilon$, $\mu$,
$\alpha$ and $\xi$. But before establishing these expressions, the
relativistic crystalline group $G'$ has to be determined. The definition of
$G'$ is the following:
$G'$ is the maximal group satisfying the relations \cite{4}
\[G' \subseteq N(G)\cap K({\bf P}_{e.m.})\]
and 
\[G = G'\cap O(3)1'\,,\]
where $G$ is the magnetic group of the crystal, $K({\bf P}_{e.m.})$ the
relativistic point group leaving invariant the polarization tensor
${\bf P}_{e.m.}$ of the crystal:
\[{\bf P}_{e.m.}\equiv {\bf P}^{\alpha\beta}_{e.m.} = \frac{1}{2}
\chi^{(\alpha\beta)(\gamma\delta)}{\bf F}_{\gamma\delta}\,,\]
and $N(G)$ ({\it i.e.} the relativistic point group associated to $G$) the
normalizer of $G$ in $O(1,3)$, {\it i.e.}:
\[N(G) = \left\{ g \in O(1,3) \, / \, g\,G\,g^{-1}
\subseteq G \right\}\,.\]
In fact in these kinds of definitions, one can also consider the
polarization tensor ${\bf P}_{e.m.}$ not only defined as the polarization
tensor of the whole crystal, but also from a more locally viewpoint, as the 
polarization tensor of an {\it ``elementary magnetic cell"\/} of the
crystal. We will consider this definition in all the further sections.
\par 
There are two cases of groups $K({\bf P}_{e.m.})$ which are particularly
interesting for the following discussion, {\it viz.}, $K_M$ and 
$K_{\angle}(a,b)$, when using Asher notations \cite{4}. These groups leave
invariant the following polarization vector $\vec P$ 
and magnetization vector $\vec M$:
\[K_M:\quad  {\vec M} (\neq \vec{0}) /\!/ O_z\quad \mbox{and}\quad \vec P =
\vec{0}\, ,\]
\smallskip
\[
K_{ \angle}(a,b):
\left\{
\begin{array}{l}
\vec M.\,\vec P\neq 0\quad\! \mbox{and}\quad\! \vec M.\,\vec P\neq 
{\parallel}{\vec M}{\parallel}.{\parallel}{\vec P}{\parallel}\, ,\\
\\
\vec P/\!/O_z\quad\! \mbox{and}\quad\! \vec M\perp O_y\,,\\
\\
\mbox{with}\,\, a=c\,\vert M_x\vert/{\parallel}{\vec P}{\parallel}\,,\\
\\ 
b = c\,\vert M_z\vert/{\parallel}{\vec P}{\parallel}\,,\\
\\
\mbox{and}\,\,  a\,b\, \neq\, 0\, .
\end{array}
\right.
\]
The generators of $K_{ M}$ and $K_{\angle}(a,b)$
are
\[K_M \equiv \left\{m'_y\,,\,\bar{1}\,,\,R_z(\phi)\,,\,L_z(\chi)\,/\,
\forall\, \phi\,,\,\chi\in\mathbb{R}\right\}\, ,\]
\[
\begin{array}{l}
K_{\angle}(a,b)\,\equiv\ \bigl\{
m'_y\,,\,L(\phi;a,b)\,,\,{\overline L}(\chi;a,b)\\ 
\qquad\qquad\qquad\qquad\qquad\quad\qquad/\, 
\forall\ \phi,\chi\ \in\mathbb{R}\bigr\}\,,
\end{array}\]
where $R_z(\phi)$ is a rotation of angle $\phi$
around the z axis, $L_z(\chi)$ is a special
transformation with velocity $\beta c$ in the z direction\footnote{\,
$\chi=\cosh^{-1}([1/(1-\beta^2)]^{1\over2}) = \cosh^{-1}(\gamma)\,.$
} 
and
\begin{equation}
L(\phi;a,b)=S^{-1}(a,b)\,R_z\,(\phi)S(a,b)\,,
\label{1}
\end{equation}
\[
{\overline L}(\chi;a,b)=S^{-1}(a,b)L_z(\chi)S(a,b)\,,
\]
where
\[
S(a,b)=
\left(
\begin{array}{rrrr}
C&0&-S&0\\
0&c&0&-s\\
-S&0&C&0\\
0&s&0&c
\end{array}
\right)
\,,
\]
with $S(a,b)\in SO_y(1,1)\times SO_y(2)$,
\[
\begin{array}{ll}
C= \cosh(\xi)\,,&S = \sinh(\xi)\,,\\
c=\cos(\omega)\,,& s = \sin(\omega)\,,
\end{array}
\]
and $\xi$, $\omega$ are functions of a and b.
Then, the procedure for establishing the constraints on the
components of the tensor $\chi$ is the following: considering that
$G=m'_y$ for example, then the normalizer $N(G)$ of $m'_y$ in $O(1,3)$
is \cite{4}
\[N(G) = O_y(1,1)\times O_y(2)\,.\]
We deduce that if $K({\bf P}_{e.m.}) = K_M$ then the generators
of the relativistic crystalline group $G'_M$ are
\[G'_M \equiv \left\{m'_y\,,\,R_z(\pi)\right\}\,,\]
whereas if $K({\bf P}_{e.m.}) = K_{\angle}(a,b)$:
\[G'_{\angle}(a,b) \equiv\left\{m'_y\,,\,L(\pi;a,b)\right\}\,.\]
The constraints on the tensor $\chi$ obviously derive from the
relations ($\forall\, h\,\in\, G'$)
\[
\chi^{(\alpha\beta)(\gamma\delta)}=
h^{\alpha}_{\mu}\,
h^{\beta}_{\nu}\,
h^{\gamma}_{\eta}\,
h^{\delta}_{\rho}\,
\chi^{(\mu\nu)(\eta\rho)}\,.
\]
Clearly, these constraints change in function of $G'$ itself, absolutly
depending on the polarization and magnetization vectors. Thus, if for
example the system passes from $\vec P=\vec{0}$ to $\vec P\neq\vec{0}$,
it involves naturally a  {\it ``breaking of the relativistic crystalline
symmetry"\/}. We must pay attention to this definition of the breaking of a
symmetry, since usually a breaking involves a reduction to a proper
subgroup of a given one, whereas in this case $G'_M$ and $G'_{\angle}(a,b)$
are conjugate groups in $O(1,3)$ (see relation (\ref{1})). But nevertheless
it can be considered as a breaking because the conjugate group is not a
subgroup and the intersection of the two is a proper one of the broken
group.

\section{\normalsize\hskip-.5cm - Relativistic broken symmetries and
toroidal moments}

We can summarize the previous results as follows.
\par 
By defining:
\[
\bar{1}\,L_y(a,b)=
S^{-1}(a,b)\,
\bar{1}\,S(a,b)\,,
\]
we have
\[
\left\{
\begin{array}{l}
S(a,b)\ \in\ N(G)\,,\\
\\
K^S_M = S(a,b)^{-1}\,K_M\,S(a,b)\\
\qquad
= K_{\angle}(a,b)\,\bar{1}\,L_y(a,b)\,,\\
\\
N(G)\,\cap\,\left\{\bar{1}\,L_y(a,b)\right\}
\ =\ \{1\}\,,\\
\\
K_{\angle}(a,b)\triangleleft K^S_M\,,
\end{array}
\right.
\]
and 
\begin{equation}
G'_M = S(a,b)\,G'_{\angle}(a,b)\,S^{-1}(a,b)\,.
\label{2}
\end{equation}
These results can be generalized in the case of a group $K_M$
corresponding to a magnetization vector $\vec M$ in the perpendicular plane
of the z-axis and not necessary parallel to $O_z$.
\par
Then, we would have ($O_M:\,\vec M$-axis)
\[
G'_M(\theta) = \{ m'_y\,,\,R_{O_M}
(\pi)\} = R_y(\theta)\,G'_M\,R^{-1}_y(\theta)\,,
\]
where $\theta$ is the angle between $\vec M$ and $O_z$.
\par
Obviously, the relation (\ref{2}) would be analogous, replacing $S(a,b)$ by
$R_y(\theta)S(a,b) = S(a,b,\theta) = S(a',b')$. It follows also from this
relation (2) that the broken symmetry we generate, is again associated to
the concept of conjugaison in the group $O(1,3)$  with respect to the
connected component of the $N(G)$ group. Then, we consider the following
classical physical system: $G'_M(\theta)$ is the relativistic crystalline
group of an elementary polarized cell of the crystal, with an electron not
in interaction with this cell in the initial state. We also assume that
this electron has initially (before intraction) a Galilean motion with a
velocity
$\vec{v}$ parallel to
$O_y$ and that the axis of motion is at the distance $x_o \neq 0$ of the
$y$-axis. Then, the electron  interacts with the cell, inducing (by a  {\it
``local magnetoelectric effect"\/}), an electric polarization, breaking the
initial relativistic crystalline symmetry group
$G'_M(\theta)$.
\par
During the interaction, the resulting symmetry group
is $G'_{\angle}(a',b')$. The transformation from one symmetry to the
other is realized with the matrix $S(a',b') \in SO_{ y}(1,1)\times
SO_y(2)$. It means that we pass with $S(a',b')$, from one frame to another
in which the motion of the electron is kept, {\it i.e.}, Galilean. In some
way, we can speak of a kind of  {\it ``inverse kinetomagnetoelectric
effect"\/} \cite{15}. Thus, the motion of the electron in the laboratory
frame is determined by $S(a',b')$, {\it i.e.}, a rotation around the
$y$-axis and a boost along the same axis. From the latter characteristics
of the motion of the electron and those of $S(a',b')$, the electron will
possess a solenoidal motion around the $y$-axis.
\par
Then, considering similarly another cell and an electron moving with a fast
solenoidal motion along an axis parallel to the $y$-axis, the interaction
with the cell might  provide a  closure of the solenoidal configuration
into a toroidal configuration in case of an adiabatic process.  Let us
precise this dynamic from a more mathematical viewpoint in the case of a
``classical motion"\/ only, since this latter is phenomenologically closed
to  the quantum one.

\section{\normalsize\hskip-.5cm - The evolution equation} 

We have precised the meaning of the transformation $S$: it equivalently
defines a change of Galilean frame during the electron-cell interaction. We
will use  this \textit{``equivalence principle"\/} (*) to obtain the type of motion or
trajectories of the  interacting electrons. We consider that passing from a
non-interacting electron-cell system to an interacting one, is equivalent
to pass from a Galilean frame ${\bf R}_t$ to another ${\bf R'}_{t'}$,  with
the transformation $S$ keeping in these two frames, the direction of a
velocity vector
$\vec v$ invariant, irrespectively of its module.
\par
The coordinates of the  position vector of the electron will be
${\widetilde{X}}=(t,x,y,z)$ in ${\bf R}_t$, and
${\widetilde{X}'}=(t',x',y',z')$ in
${\bf R'}_{t'}$.
\par It follows that
\begin{equation}
\widetilde{X}' = S\,\widetilde{X}\,,
\label{3}
\end{equation}
where $S\in N(G)$ and $S$ function of $t'$ (or $t$).
\par
At this stage, if $\widetilde{X}$ and $\widetilde{X}'$ have a variation due to
an infinitesimal change of the ${\bf R}_t$ and ${\bf R'}_{t'}$  frames, then
in order to keep the relativistic covariance of the relation (\ref{3}), this
transformation has  to be an infinitesimal Lorentz transformation in
$N(G)$. Hence
\[\widetilde{X}' + d\widetilde{X}' = ({\bf 1} - dn')
\widetilde{X}'\,,\]
\begin{equation}
\widetilde{X} + d\widetilde{X} = ({\bf 1} - dn)
\widetilde{X}\,,
\label{4}
\end{equation}
where $dn$ and $dn'$ are elements of the Lie algebra of $N(G)$.
Then, from equation (\ref{3}), we deduce that
\[d\widetilde{X}' = (dS)\widetilde{X}\ +\ S\,d\widetilde{X}
= ((dS){S^{-1}})\widetilde{X}'
+ S\,d\widetilde{X}\,,\]
and from equation (\ref{4})
\[
d\widetilde{X}' = ((dS)S^{-1})\widetilde{X}' -
(S\,dn\,S^{-1})\widetilde{X}'\,,\]
\begin{equation}
d\widetilde{X}' = ((dS){S^{-1}} -
S\,dn\,{S^{-1}})\widetilde{X}' = -dn'\widetilde{X}'\,.
\label{5}
\end{equation}
Now, if we ascribe the variations of $\widetilde{X}$ and $\widetilde{X}'$ to 
the determination of these two 4-vectors in a third frame ${\bf
R''}_{t''}\equiv{\bf R'}_{t'+dt'}$  obtained from ${\bf R'}_{t'}$ by an
infinitesimal evolution in time:
$dt'$, of this latter frame, then  equation (\ref{5}) can be written as
\[
\frac{d\widetilde { X}'}{dt'} = \left(\left(\frac{dS}{dt'}\right)S^{-1} -
S\,\left(\frac{dn}{dt'}\right)\, S^{-1}\right)\widetilde{X}'\,.\]
Of course, we recognize a gauge transformation with respect to the
$N(G)$ group of an  affine connexion $dn/dt'$ of the frames bundle of
the Minkowski space, since we have the well-known relation in gauge theory
\[dn' = S\,dn\,S^{-1} - (dS)\,S^{-1}\,.\] 
More, we recognize particularly a kind of {\it Thomas precession  equation}
\cite{17,18} which explains the solenoidal motion of the electron  during
the interaction with the magnetic cell. Let us add that in accordance with
the equivalence principle (*) we gave, the ${\bf R''}_{t''}$ frame can be
associated  to an evolution of the electron-cell interaction (or of the
breaking of the  relativisic crystalline symmetry of the cell). If for
instance,  $S$ is of the form
\[\footnotesize
\begin{array}{l}
S=\\
\left(
\begin{array}{cccc}
\cosh(\xi t')&0&-\sinh(\xi t')&0\\
0&\cos(\omega t')&0&-\sin(\omega t')\\
-\sinh(\xi t')&0&\cosh(\xi t')&0\\
0&\sin(\omega t')&0&\cos(\omega t')
\end{array}
\right)
\end{array}
\]
and $\widetilde{X}= (ct,x_o,vt,0)$, where $v={\parallel}{\vec v}{\parallel}$
is the module of the invariant velocity vector $\vec v$, we obtain
\[
\left\{
\begin{array}{rcl}
ct'&=& ct\cosh(\xi t') - vt\sinh(\xi t')\,,\\
x'(t')&=& x_o\cos(\omega t')\,,\\
y'(t')&=&vt\cosh(\xi t') - ct\sinh(\xi t')\,,\\
z'(t')&=&x_o\sin(\omega t')\,.
\end{array}
\right.
\] 
Now, because of the equivalence principle (*), the relations above must be 
satisfied whatever is the module $v$. Then, we have necessarily $\xi=0$ and
the time
$t'$ has to be equaling to the time
$t$ of the laboratory frame. Then $y'(t) = vt$ and $x'(t)^2+z'(t)^2=1$.
Hence, the electron has a solenoidal motion as expected. In conclusion,
only $SO(2)$ acts on the dynamic of the electron because of the equivalence
between $t$ and $t'$, whereas $N(G)$ is the gauge group of the
electron-cell  system. To conclude, the motion of the electron is invariant
with respect to the maximal group contained in
$N(G)\cap O(3)$. Let us remark that the inclusion in $O(3)$ is a well-known
result of the precession theory.
\par 
In the case of a polarized electron, the Thomas precession of its spin
along the trajectory would lead to a toroidal configuration of spin
currents \cite{16}. On the other hand, if the electron has a fast solenoidal
motion before the  interaction with the cell, then with the condition that
the interaction is adiabatic, one will obtain, by the latter  process, a
slow solenoidal motion of the mean position of the electron, and so a
toroidal motion during the interaction.

\section{\normalsize\hskip-.5cm - The compatible Shubnikov groups}

In the case of the $m'_y$ symmetry, such toroidal motions cannot be
completly  compatible with this symmetry because they produce a
non-vanishing orbital moment along the $y$-axis. Then, in order to have a
non-zero toroidal moment, it is necessary to  assume the existence of a
second electron coupled with the first one with the same toroidal moment but
with an opposed orbital moment. In this case the crystal will be in a kind
of ``ferro-toroidal" phase ($T^2$) (the exponent 2 for two electrons).
\par
One can generalize the latter procedure to other magnetic groups. The
groups $G$ for which toroidal moments may exist, must possess a normalizer
$N(G)$ containing the group $O(2)$. From the classification of the
normalizers given by E. Asher, one can state  that the only (fifteen groups
plus the group 1) compatible magnetic  groups are \cite{4}
\[\small
\begin{array}{cccccccc}
1,&\bar{1}',&2,&2',&3,&3',&4,&4',\\
6,&6',&m,&m',&2/m',&2'/m,&4/m',&6/m'.
\end{array}
\]
Then, applying the latter tools, we can work out the toroidal phases for
each magnetic group, knowing the possible orientations of the magnetic,
electric and kinetic polarization and the permitted 
$K({\bf P}_{e.m.})$ groups. The results are given in Table I, following the 
H. Schmid classification and notations \cite{6}.
\par
In order to illustrate how to construct the Table I, one can give two
examples: the group $m$, and a AA-type one. In the case of $m$, the 
orientations of $\vec P$, $\vec M$ and $\vec v$ are represented in figure 3.
The normalizer $N(m)$ is: $O_{xz}(1,2){\dot \times}m_{ y}$. The
possible  basic continuous transformations from one polarization state to
another one are ($S(a)$: boost, $R_y(\theta)$: rotation around the
$y$-axis with an angle $\theta$. The null polarization and magnetization
vectors are not  indicated):
\[
\left\{
\begin{array}{rcl}
\vec P&\,\longrightarrow\,&\vec P'\quad\mbox{and}\quad\widehat{(\vec
P,\vec P')}  =\theta\,,\\
\\
K_P&\,\longrightarrow\,&
K_{\perp}(a,\theta) =\\
&&\, R^{-1}_y
(\theta)S^{-1}(a)K_PS(a)R_y(\theta)\,,
\end{array}
\right.\hspace{3mm}
\]
\[
\left\{
\begin{array}{rcl}
\vec M&\,\longrightarrow\,&\vec P\quad\mbox{and}\quad \vec M'/\!/\vec M\,,\\
\\
K_M&\,\longrightarrow\,&K_{\perp}(a) = S^{-1}(a)K_M
S(a)\,,
\end{array}
\right.\hspace{6mm}
\]
\[
\left\{
\begin{array}{rcl}
\vec M&\,\longrightarrow\,&\vec M'/\!/\vec M\,,\\
\\
K_M&\,\longrightarrow\,&K_{M'} = K_M\,,
\end{array}
\right.\hspace{30mm}
\]
\[
\left\{
\begin{array}{rcl}
\vec P&\,\longrightarrow\,&\vec P' \quad\mbox{and}\quad
\widehat{(\vec P,\vec P')}  = \theta\,,\\
\\
K_P&\,\longrightarrow\,&K_{P'}(\theta) = R^{-1}_y(\theta)K_PR_y(\theta)\,.
\end{array}
\right.\hspace{6mm}
\]
All the other transformations can be deduced from these four. The
computation of the groups $G'$ in each case shows that no conjugaisons are
possible between two groups $G'$ with a rotation around an axis contained
in the plane $m$. It follows that no broken symmetry appears
and consequently no toroidal moments contained in the plane $m$. Only
conjugaisons with $R_y(\theta)$ can be taken, which may suggest the
existence of kind of  ``antiferro-toroidal phases" along the $y$-axis 
($\overline{T}_y$) (Two opposite toroidal moments per magnetic cell along
the $y$-axis). In the case of $G$ groups of type AA, any orientations in
each sublattices of the crystal are permitted so that toroidal moments may
appear if oblique fields exist since in this case, it is always possible to
conjugate by a rotation two $K({\bf P}_{e.m.})$ groups. As a result, making
the same kind of computation as above, we find that only 12 groups among the
16 ones tabulated in Table I are associated to a non-trivial $SO(2)$ action,
and then can induced a creation of a solenoidal motion, \textit{i.e.}
anyons.
\par
From these results, one can do few hypothesis as conclusion about a model
of superconductivity in magnetoelectric crystals.

\section{\normalsize\hskip-.5cm - Conclusion}

We have seen in the previous sections that the breaking of relativistic
symmetry can produce a toroidal moment. But in the same time, the electron,
interacting with the cell, acquires an orbital moment, whereas, in the
initial state, the orbital moment of the electron-cell system was zero. As
a  result, a quasi-particle is produced during the interaction with a
non-zero orbital moment in order to keep the total orbital moment. This
quasi-particle is associated with a broken symmetry. Strictly speaking, it
is a Goldstone boson. The latter one is connected with the propagation in
the crystal of the broken symmetry, {\it i.e.}, the variation of the local
polarization and magnetization resulting from a magnetoelectric interaction
between cells and electrons. Consequently, it will carried an electric and
magnetic moment and  a non-vanishing orbital moment.
\par 
Then, one can consider a superconductivity model fitting the B.C.S.
theory where the Goldstone bosons, and not the phonons, will couple the
electrons of the Cooper pairs. Models of the electron-hole pairing
involving  toroidal moments have been already studied in the case of
excitonic-insulators \cite{13}.  Let us mention that the Goldstone bosons
will interact between themselves via their electric and magnetic moments,
{\it i.e.}, via segnetomagnon modes. Obviously, such  bosons can be directly
responsible for the coupling of two toroidal moments in the $T^2$
phase for example.
\par
We can precise the characteristics of the Goldstone bosons considering the
cyon theory \cite{19,20,23}. Indeed, the cyons are types of atoms with one
electron and no nucleus in which the electron is constrained to move in a
toroidal domain of the space. The cyon theory has been linked with  another
one: the dyon theory \cite{19,20,23}. The dyons are made of one electron and
a Dirac monopole as nucleus of the atom. In the model described in this
paper, in order to explain the origin of toroidal moments, we considered an
electron with a solenoidal motion in the initial state.
\par
Bearing in mind that the potential vector $\widetilde{A}$ of a Dirac monopole
can be generated by a semi-infinite solenoid of electrical currents along
the semi-infinite  singularity axis of  the potential vector, we can
consider that the interaction of the electron with the cell will produce a
dyon in which the single electron would have an orbital with a toroidal
configuration. Hence, we would obtain a cyon and a Dirac monopole. Let us
note that a link exists between toroidal moments and magnetic charges, as
it has been already shown \cite{13}.  The toroidal configuration will be
associated with an invariance of the electron-cell system  with respect to
the group $N(G)$ only (containing $SO(2)$ as a gauge group of the dyon and
of  the potential vector of the monopole). Then, it follows that we may
consider the Goldstone bosons associated with the breaking of the
relativistic symmetry as an effective Dirac monopole. We can also precise
that we assumed an adiabatic electron-cell interaction so that a Berry phase
appears from a Yang-Mills gauge field in the equivalent quantum system. It
has been shown that  this gauge field can be associated to an anomaly such
as a monopole for example.
\par
If we turn out to Lagrangian formalism, the latter remark means that
beginning with a Lagrangian describing the electrons and invariant with
respect to the space group of the crystal, we will consider the interaction
with the cell in the crystal via magnetoelectric effects, adding
perturbative Lagrangians coming from a gauge interaction associated to the
gauge group $N(G)$. This group will break the relativistic crystalline
symmetry of the cell during the interaction. The resulting electromagnetic
gauge field will be  associated to the effective  Dirac monopole in such a
way that the total Lagrangian be invariant only with respect to the group
$N(G)$ (the connected component of $N(G)$ being the gauge group).
Furthermore, the question arises to know if two electrons with a toroidal
configuration of currents, and consequently producing a toroidal moment,
may described anyons and the breakdown of time-reversal (T) and parity (P) 
symmetries
\cite{21,22}. Indeed, toroidal moments break these two symmetries but not
the PT one as expected in anyon theory. More, cyons are anyons, and the flux
tube of anyons can be generated by a monopole \cite{23}.
\par
From an experimental viewpoint, it would be necessary to characterize the
toroidal moments by interaction with light or hyperfrequency electric
or/and magnetic fields. Absorption of far infra-red light under static
magnetic fields have been applied and measurements of the circular 
dichroism are now available \cite{9,13}. To close, let us note  that under
highfrequency fields the crystal would present a diamagnetic-paramagnetic
transition \cite{13}.
\par\bigskip
\noindent\textbf{\Large Acknowledgements}\par\medskip
The author greatly thanks Prof. H. Schmid of the
\textit{University of Geneva\/} who suggested the present work. I have appreciated his
great experience on this subject and his hospitality in his laboratory. I
would like also to thank Dr. J.-P. Rivera of the \textit{University  of Geneva\/} for
his relevant remarks. I would like also to dedicate this work to  Prof. M.
Kibler of the {\it Institut de Physique Nucl\'eaire de l'Universit\'e
Claude Bernard} (Lyon, France) for his continuous help and encouragements.

\bigskip
\noindent\textbf{\Large Figure captions}\par\medskip

\noindent \underline{\bf Table I:}
\begin{itemize}
\item[$\odot$ :] ``Weak ferromagnetism" permitted (I. F. Dzyaloshinskii, (1957)
\cite{14}).
\item[$\dagger$ :] ``Weak ferroelectricity" permitted.
\end{itemize}
\bigskip

\noindent \textsc{Type of ordering:}
\begin{itemize}
\item  $M$ = pyro-, ferro- or ferrimagnetic;\\ $P$ = pyro-, ferro-
or ferrielectric.
\item $\overline{M}$ = antiferromagnetic;\\ $\overline{P}$ =
antiferroelectric or orthoelectric.
\item $T$ = pyro-, ferro-, or ferritoroidal;\\ $T^2$ = same
phases as $T$ but with two electrons per toroidal moment.
\item $\overline{T}$ = antiferrotoroidal phase;\\ $\overline{T}^2$ = same
phases as $\overline{T}$ but with two  electrons per toroidal moment.
\item $\overline{T}_y$ = antiferrotoroidal phase along the
$y$-axis.
\item H: spontaneous magnetization permitted;\\ E: spontaneous polarization
permitted.
\item EH: linear magnetoelectric effect permitted.
\item EEH and HEE: second-order magnetoelectric effect.
\item $P^S$: invariant electric polarization vector; $M^S$: invariant
magnetization vector;\\ $T^S$: invariant toroidal vector;\\
$v$: velocity vector.
\item +: meaning all directions permitted.
\item ${\dot \times}$: semi-direct product.
\item ($y$): perpendicular to the $y$-axis, $y$: along the $y$-axis.
\item $m_v$: symmetry with respect to  the $v$-plane.
\end{itemize}
\bigskip

\noindent \underline{\bf Figure 2:}\\
Configuration of a current having a toroidal dipole moment. The arrows 
on the torus indicate the current direction, and the moment is directed along
the symmetry axis of the torus.\\
\bigskip

\noindent \underline{\bf Figure 3:}\\
\noindent Orientations of the polarization vectors $P$ and $M$, and 
the velocity vector $v$ in case of a $m$ symmetry.

\vfill\eject
\vfill
\noindent\includegraphics[bb=142 0 418 790,scale=.83]{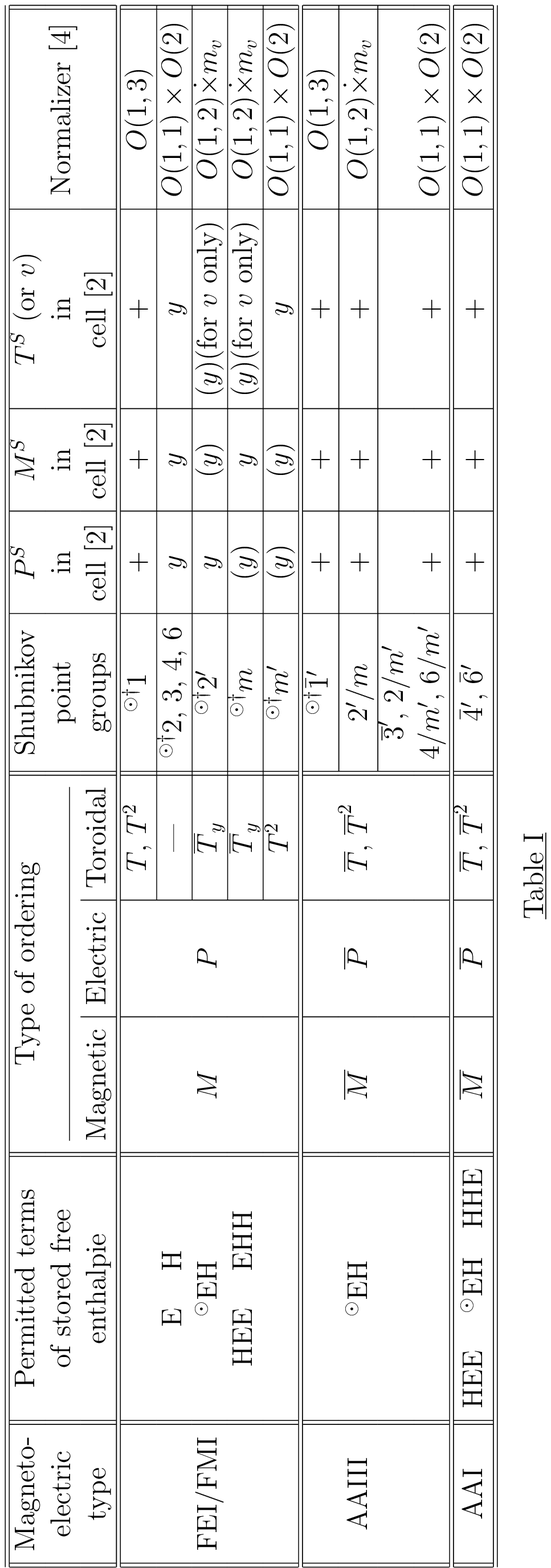}
\vfill\eject
\mbox{}\vfill
\hskip-1.4cm\fbox{\includegraphics[bb=19 3 563
716,scale=.85]{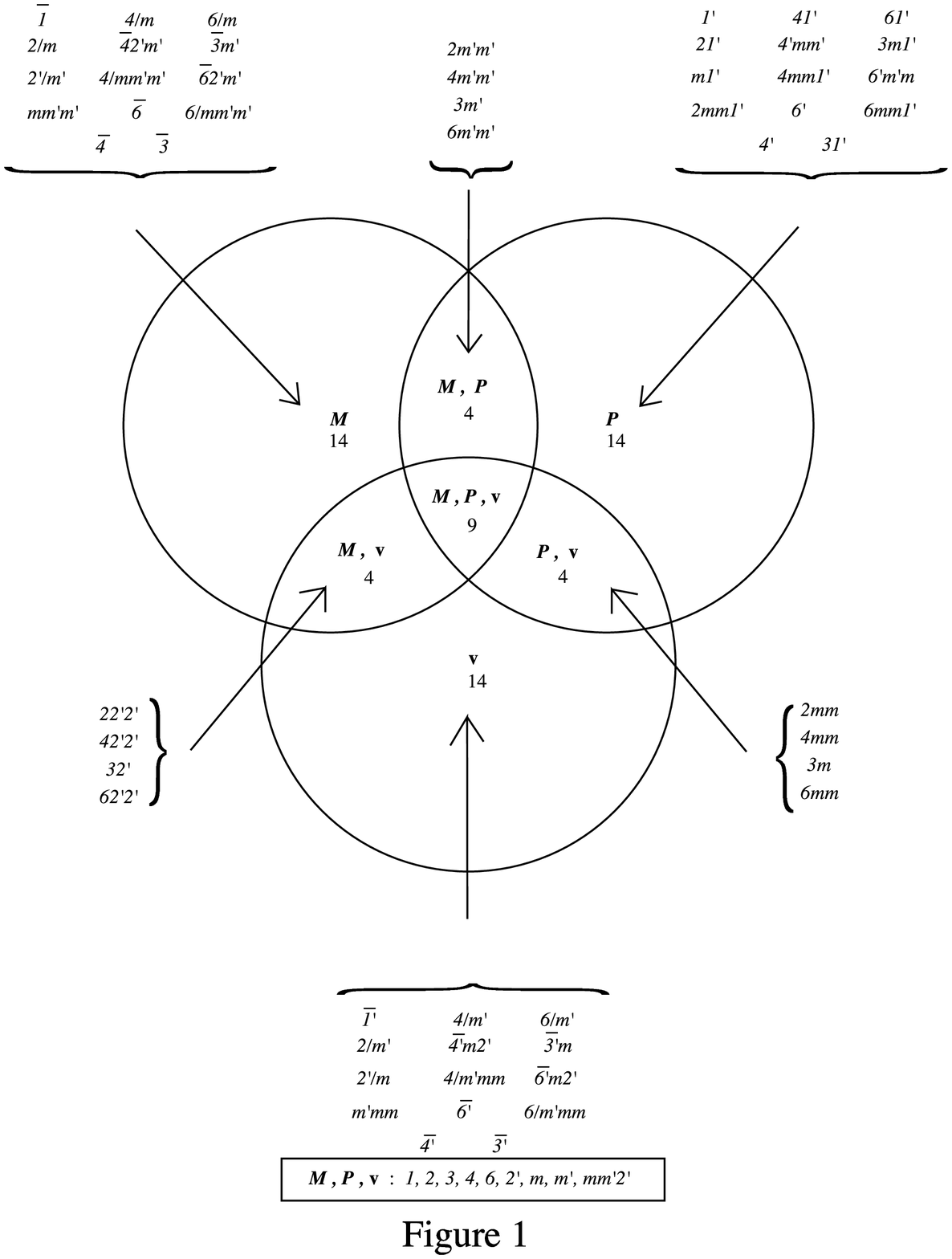}}
\vfill\eject
\mbox{}
\vfill\eject
\vfill
\hskip1cm\fbox{\includegraphics[bb=89 169 473 811,scale=1]{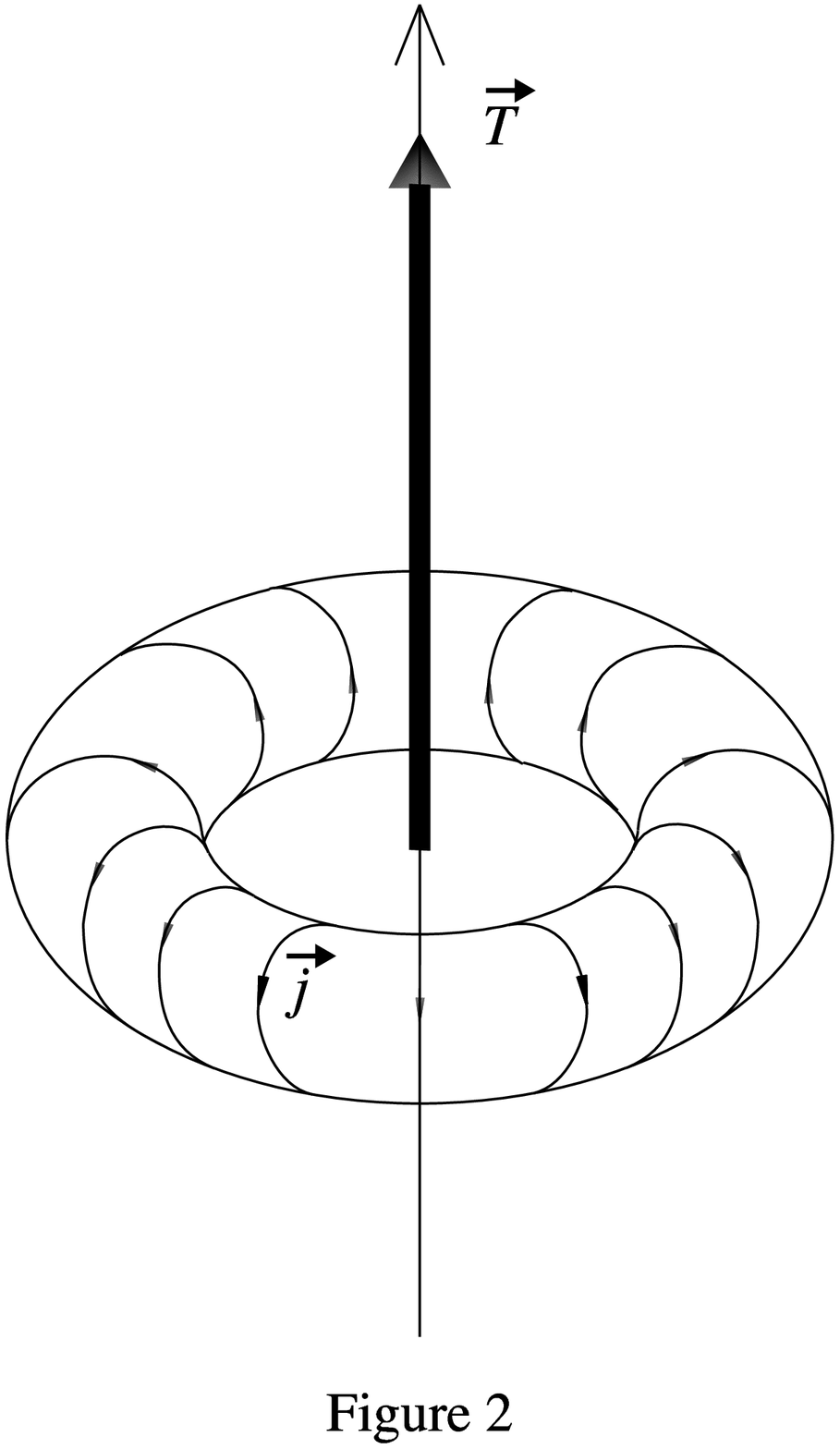}} 
\vfill\eject 
\mbox{}
\vfill\eject
\vfill
\hskip-1cm\fbox{\includegraphics[bb=42 144 488
700,scale=1]{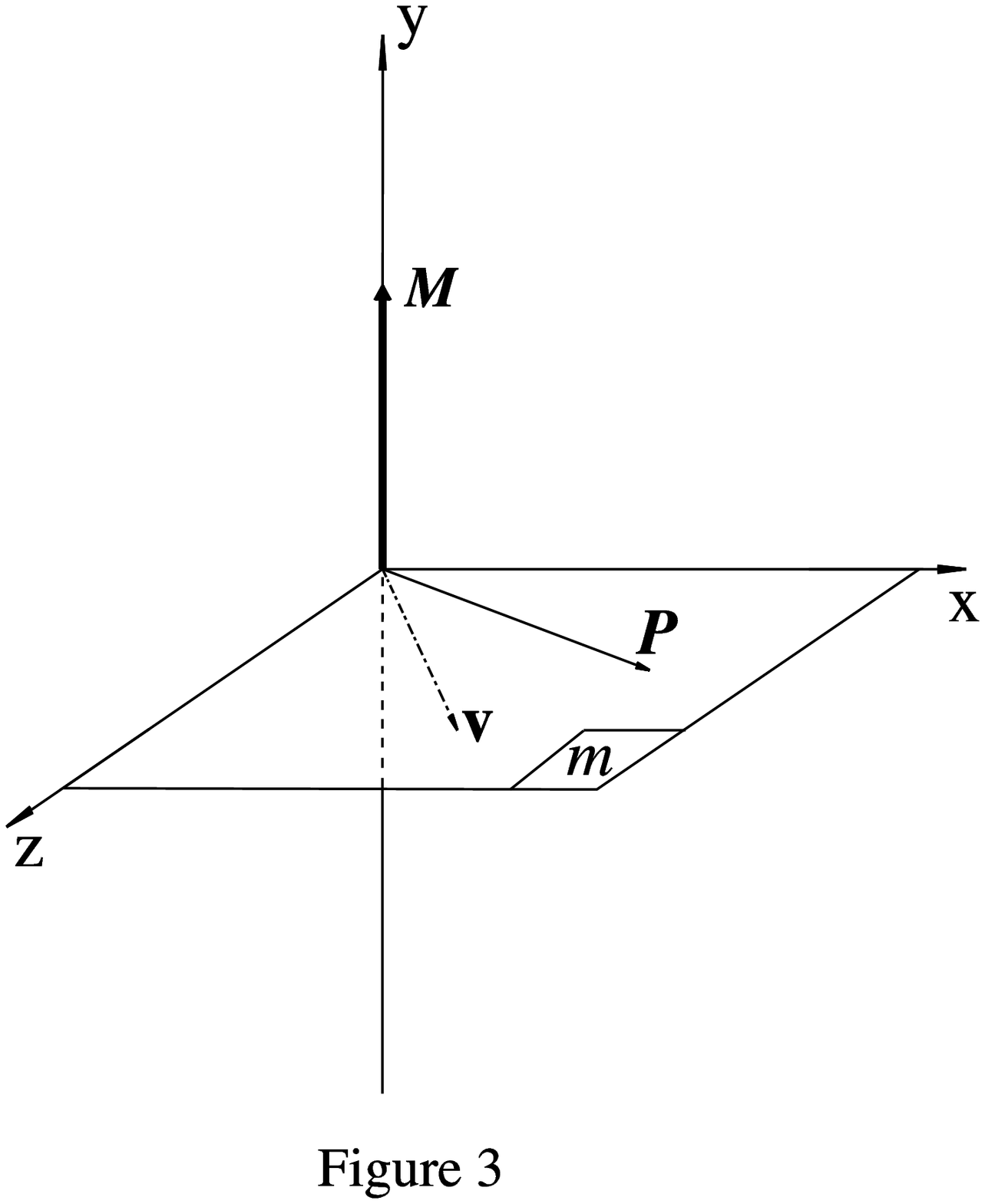}} 
\vfill\eject
 
\end{document}